\input harvmac.tex
\let\includefigures=\iffalse
\let\useblackboard=\iffalse
\def\Title#1#2{\rightline{#1}
\ifx\answ\bigans\nopagenumbers\pageno0\vskip1in%
\baselineskip 15pt plus 1pt minus 1pt
\else
\def\listrefs{\footatend\vskip 1in\immediate\closeout\rfile\writestoppt
\baselineskip=14pt\centerline{{\bf References}}\bigskip{\frenchspacing%
\parindent=20pt\escapechar=` \input
refs.tmp\vfill\eject}\nonfrenchspacing}
\pageno1\vskip.8in\fi \centerline{\titlefont #2}\vskip .5in}

\ifx\answ\bigans\def\tcbreak#1{}\else\def\tcbreak#1{\cr&{#1}}\fi
%

%
\def\yboxit#1#2{\vbox{\hrule height #1 \hbox{\vrule width #1
\vbox{#2}\vrule width #1 }\hrule height #1 }}
\def\fillbox#1{\hbox to #1{\vbox to #1{\vfil}\hfil}}
\def\ybox{{\lower 1.3pt \yboxit{0.4pt}{\fillbox{8pt}}\hskip-0.2pt}}
\def\l{\left}
\def\r{\right}
\def\comments#1{}

\def\p{\partial}

\def\eps{\epsilon}

\def\tr{{\rm tr\ }}

\def\ket#1{|#1\rangle}
\def\vev#1{\langle{#1}\rangle}

\def\CN{{\cal N}}

\def\CL{{\cal L}}

\def\nl{\hfill\break}

\def\II{\relax{I\kern-.10em I}}
\def\IIa{{\II}a}

\def\IZ{\relax\ifmmode\mathchoice
{\hbox{\cmss Z\kern-.4em Z}}{\hbox{\cmss Z\kern-.4em Z}}
{\lower.9pt\hbox{\cmsss Z\kern-.4em Z}}
{\lower1.2pt\hbox{\cmsss Z\kern-.4em Z}}\else{\cmss Z\kern-.4em
Z}\fi}
\def\IB{\relax{\rm I\kern-.18em B}}
\def\IC{{\relax\hbox{$\inbar\kern-.3em{\rm C}$}}}
\def\ID{\relax{\rm I\kern-.18em D}}
\def\IE{\relax{\rm I\kern-.18em E}}
\def\IF{\relax{\rm I\kern-.18em F}}
\def\IG{\relax\hbox{$\inbar\kern-.3em{\rm G}$}}
\def\IGa{\relax\hbox{${\rm I}\kern-.18em\Gamma$}}
\def\IH{\relax{\rm I\kern-.18em H}}
\def\II{\relax{\rm I\kern-.18em I}}
\def\IK{\relax{\rm I\kern-.18em K}}
\def\IP{\relax{\rm I\kern-.18em P}}

%

\def\p{\partial}

\font\cmss=cmss10 \font\cmsss=cmss10 at 7pt
\def\IR{\relax{\rm I\kern-.18em R}}

\def\drho{{\dot\rho}}

\def\inbar{\,\vrule height1.5ex width.4pt depth0pt}

\def\lp10{l_P^{10}}
\def\lp11{l_P^{11}}
\def\R11{R_{11}}

\Title{\vbox{\baselineskip14pt\hbox{hep-th/9610236}
\hbox{RU-96-100}}}
{\vbox{
\centerline{Five-branes in M(atrix) Theory} }}
\centerline{Micha Berkooz and Michael R. Douglas}
\medskip
\centerline{\it Department of Physics and Astronomy}
\centerline{\it Rutgers University }
\centerline{\it Piscataway, NJ 08855--0849}
\medskip
\centerline{\tt berkooz, mrd@physics.rutgers.edu}
\medskip
\bigskip
\noindent
We propose a construction of five-branes which fill both light-cone dimensions
in Banks, Fischler, Shenker and Susskind's matrix model of M theory.
We argue that they have the correct long-range fields
and spectrum of excitations.
We prove Dirac charge quantization with the membrane by
showing that the five-brane induces a Berry phase in
the membrane world-volume theory, with a familiar magnetic monopole form.
\Date{October 1996}

\def\laplace{{\kern1pt\vbox{\hrule height 1.2pt\hbox{\vrule width 1.2pt\hskip
  3pt\vbox{\vskip 6pt}\hskip 3pt\vrule width 0.6pt}\hrule height 0.6pt}
  \kern1pt}}
\def\scriptlap{{\kern1pt\vbox{\hrule height 0.8pt\hbox{\vrule width 0.8pt
  \hskip2pt\vbox{\vskip 4pt}\hskip 2pt\vrule width 0.4pt}\hrule height 0.4pt}
  \kern1pt}}

\lref\BFSS{T. Banks, W. Fischler, S. H. Shenker and L. Susskind,
``M Theory As A Matrix Model: A Conjecture,'' hep-th/9610043.}
\lref\dewit{B. de Wit, J. Hoppe and H. Nicolai, Nucl.~Phys. B305(1988)
 545.\nl
B. De Wit, M. Luscher and H. Nicolai, Nucl.~Phys. B320(1989) 135.}
\lref\dzero{U.H. Danielsson, G. Ferretti and  B. Sundburg,
``D-particle Dynamics and Bound States'', hep-th/9603081\nl
D. Kabat and P. Pouliot, ``A Comment on Zero-Brane Quantum
Mechanics'', hep-th/9603127}
\lref\DKPS{M. R. Douglas, D. Kabat, P. Pouliot and S. Shenker,
``D-branes and Short Distances in String Theory,'' hep-th/9608024.}
\lref\Berry{M. V. Berry, Proc. Roy. Soc. London {\bf A392}, 45;
reprinted in A. Shapere and F. Wilczek, {\it Geometric Phases in Physics},
World Scientific 1989.}
\lref\shenker{S.~H.~Shenker, ``Another Length Scale in String Theory?,''
hep-th/9509132.}
\lref\PolRR{J.~Polchinski, Phys.~Rev.~Lett.~{\bf 75} (1995) 4724,
hep-th/9510017.}
\lref\joerev{S.~Chaudhuri, C.~Johnson, and J.~Polchinski,
{\it Notes on D-Branes}, hep-th/9602052.}
\lref\kp{D.~Kabat and P.~Pouliot,
{\it A Comment on Zero-Brane Quantum Mechanics}, hep-th/9603127.}
\lref\guven{R. Gueven, Phys. Lett. B276 (1992) 49.}
\lref\duff{M. J. Duff, hep-th/9608117.}
\lref\sethi{S.~Sethi and M.~Stern, ``A Comment
on the Spectrum of H-Monopoles,'' hep-th/9607145.}
\lref\nepteit{R.~Nepomechie, Phys. Rev. {\bf D31} (1985) 1921;\nl
C.~Teitelboim, Phys. Lett. {\bf B167} (1986) 63, 69.}
\lref\naka{M. Nakahara, {\it Geometry, Topology and Physics,} Adam
Hilger, 1990; pp. 369-371.}
\lref\shenap{S. H. Shenker, private communication and to appear
in a revised version of \BFSS.}
\lref\toappear{O. Aharony, M. Berkooz, M. R. Douglas, S. H. Shenker
and others,
work in progress.}

\newsec{Introduction}

Recently Banks et. al. have proposed a definition of
eleven-dimensional M theory in the infinite momentum frame \BFSS,
as a large $N$ limit of maximally supersymmetric
matrix quantum mechanics.  This system has a rather unusual history --
it was first studied as a regulated supermembrane
theory \dewit, and later arose as the theory governing the short distance
dynamics of D$0$-branes in type \IIa\ superstring theory \dzero.
The results following from both studies fit naturally into
their picture.

In this note we propose a definition of certain five-branes
in this theory, and check a number of the known properties of the
five-brane in M theory -- in particular, both the particles in the
supergravity multiplet
and the supermembrane as defined in \BFSS\ see the correct long-distance metric
and quantized magnetic four-form field strength.
This also confirms
that the supermembrane couples to the three-form gauge potential,
which is not manifest in their definition.

The five-branes we discuss are those whose world-volume includes
both light cone coordinates, or ``longitudinal five-branes.''
These are objects with zero longitudinal momentum in the ground state
and thus should be considered as non-trivial backgrounds in the IMF.
On general grounds, such a background should
correspond to a modified Lagrangian.

The specific modification we propose is inspired by the analogous system
in the type \IIa\ string, interacting Dirichlet $4$-branes and $0$-branes.
The $4$-brane arises as an M theory five-brane wrapped around the eleventh
dimension, and thus it has Kaluza-Klein excitations.  These reduce to \IIa\
$0$--$4$ bound states, and one was exhibited in \DKPS.
The short distance
interaction between these objects and thus the existence of the
bound state is entirely due to stretched open strings between the
$0$-branes and $4$-branes.  The lightest such strings
form a hypermultiplet in the vector representation of the zero-brane
gauge symmetry group.

Keeping only these lightest modes as
additional M theory degrees of freedom leads to a theory with (generically)
no new massless degrees of freedom, but a modified dynamics for the
zero-branes and new bound states.
The discussion of the five-brane Kaluza-Klein modes
is very analogous to the discussion of supergravity KK modes in \BFSS,
and the arguments they give for re-interpreting zero-brane
bound state dynamics as
the IMF dynamics of particles in the supergravity multiplet
apply here, allowing us to re-interpret the dynamics of the new
zero-brane bound states
as the IMF dynamics of particles on the five-brane world-volume, forming
a tensor multiplet.

Thus, the addition of a five-brane to the background will be implemented by
adding a hypermultiplet in the vector representation to the
Lagrangian.
The location and orientation of the five-brane will be
encoded in the couplings and Lorentz properties of the additional fields.
By integrating out this hypermultiplet, we will derive both the world-volume
spectrum and the long-range fields, as seen by particles and by membranes.

We propose the Lagrangian in section 2
and show that supersymmetry acts properly.
In section 3 we consider the interaction with $0$-branes, arguing both
that the bound states exist and that the long range fields are correct.
In section 4 we consider the membrane, and show that its global interaction
with the five-brane magnetic field can be understood as a Berry phase for
fermion zero modes on its world-volume.  This leads to a very simple
proof that it and the five-brane satisfy the Dirac condition with
the minimal charge quantum.
Section 5 contains conclusions.

\newsec{Matrix-vector quantum mechanics}

The model of \BFSS\ is a $U(N)$ gauged matrix supersymmetric
quantum mechanics.
The Lagrangian (eqn. 4.2 of \BFSS) is
\eqn\lagmat{
L= tr \biggl[ {1 \over 2R} D_t{X^i} D_t{X^i}
- \bar{\theta} \gamma_- D_t{\theta} - R\ \bar{\theta}
\gamma_- \gamma_i [\theta ,X^i]
- {1\over 4} R\ [X^i,X^j]^2  \biggr]
}
with $D_t = \p_t + iA$ and $\gamma_+\theta=0$.
The SUSY transformation laws are
\eqn\sutrans{\eqalign{
\delta_0 X^i &= - 2\bar{\eta} \gamma^i \theta \cr
\delta_0 \theta &= \ha \gamma_+ \biggl[ D_t X^i \gamma_i +
\gamma_- + \ha [X^i \ ,X^j ]\ \gamma_{ij} \biggr] \eta \cr
\delta_0 A &= - 2\bar{\eta} \theta \cr
}}
(written $\delta_0$ to distinguish them from the supersymmetries of the
Lagrangian with the five-brane added).
The $16$ supersymmetries $\tilde\eta$
satisfying $\gamma_+\tilde\eta = 0$ are realized
trivially, while the others $\gamma_-\eta = 0$ anticommute to
the Hamiltonian
\eqn\hammat{
H=R\ tr \left\{{{{ \Pi_i \Pi_i}\over 2}
+{1\over 4}\ [X^i,X^j]^2}+\bar{\theta}\gamma_- \gamma_i [\theta,X^i]
\right\}.}

In the eleven-dimensional interpretation, this is the light-cone Hamiltonian
$P_+ = H$, while the longitudinal momentum $P_-=N/R$.
$R$ is an IR cutoff on the longitudinal momentum, and both $R$ and $N$
are taken to infinity to get eleven-dimensional physics.
A state containing free massless particles indexed by $i$
each with momenta $(p_{-i},\vec p_i)$ and spin (really, element within
the supermultiplet) $s_i$
has
\eqn\disper{
P_+ = \sum_i {(\vec p_i)^2\over 2p_{-i}}
 \equiv R\sum_i {(\vec p_i)^2\over 2n_{i}}.
}
The corresponding wave function is approximately a product
\eqn\wavefn{
\Psi = \prod_i e^{i\vec p_i \tr\vec X^{(n_i)}/n_i}\
\psi_{n_i,s_i}(\vec X^{(n_i)}, \theta^{(n_i)})
}
where $\vec X = \oplus_i \vec X^{(n_i)}$ is a direct sum over factors
with rank $n_i$.  It is strongly believed that there is a unique zero
energy eigenstate $\psi_{n,s}$ for each $n\ge 1$ and $s$, so these
wave functions are in one to one correspondance with asymptotic particle
states.

A longitudinal five-brane is embedded in a four-dimensional hyperplane
in the nine transverse dimensions.  Let us take its coordinates to be
$X^m$ with $1\le m\le 4$, and the transverse coordinates to be
$X^a$ with $5\le a\le 9$.  The manifest $SO(9)$ symmetry is broken
to $SO(4)_{||} \times SO(5)_\bot$.
The hyperplane will be $X^a=x^a_0$.
Let $\rho$ and $\drho$ index the two spinor representations of
$SO(4)_{||}$ and $\alpha$ index a spinor of $SO(5)_\bot$, all
raised and lowered with antisymmetric $\eps$ symbols.
A nine-dimensional spinor reduces to a ``symplectically real''
spinor satisfying $\eta^*=\eps\eps \eta$.

The five-brane breaks half of the original $32$ supersymmetries,
$\eta^{\drho}_{\alpha}$ and $\tilde\eta^\rho_{\alpha}$,
and leaves unbroken
$\eta^\rho_{\alpha}$ and $\tilde\eta^{\drho}_{\alpha}$.
Thus it has a $2^8$-fold multiplet of ground states forming a representation of
the broken supersymmetries.  This will be represented by adding fermionic
couplings $\theta_0$ to the Lagrangian, which
transform inhomogeneously under the broken supersymmetry
$\tilde\eta^\rho_{\alpha}$.

We propose to describe the five-brane by adding a complex
boson $v^{\rho\drho}$ and fermion $\chi^{\alpha\drho}$,
both vectors under $U(N)$ and symplectically real.
The Lagrangian is
\eqn\lagvec{\eqalign{
L_5 &= |D_t{v^{\rho\drho}}|^2 + \chi D_t \chi
- v_{\rho\drho}(X^a-x^a_0)^2 v^{\rho\drho}
- {\chi}_\alpha^\drho (X^a-x^a_0)
	\gamma_a^{\alpha\beta} {\chi_{\beta\drho}} \cr
&- v_{\rho\drho} (\theta-\theta_0)_\alpha^\rho \chi^{\alpha\drho}
+ v_{\rho\drho} [X^m,X^n] \sigma_{mn}^{\rho\sigma} v_\sigma^\drho - |v|^4 .
}}
This is modeled after the dimensional reduction of $\CN=1$, $d=6$ gauge
theory and the last two terms here combine with the last term of \lagmat\
to form the usual D-terms of that theory.
The mass term could be rotated to a conventional $d=6$ mass term, but
we choose to keep $SO(5)_\bot$ manifest.

The supersymmetries now act as \sutrans\ on $X$ and
\eqn\supvec{\eqalign{
\delta v^{\rho\drho} &=
 \eta^\rho_{\alpha} \chi^{\drho\alpha} \cr
\delta \chi^{\alpha\drho} &=
  D_t v^{\rho\drho} \eta_\rho^\alpha \cr
\delta \theta^\rho_\alpha &= \delta_0 \theta^\rho_\alpha +
v^{\drho(\rho} v^{\sigma)}_\drho \eta_{\sigma\alpha}  \cr
\delta \theta^{\rho}_{0\alpha} &= \tilde\eta^{\rho}_{\alpha} \cr
\delta x_0^a &=
	- 2\eta^{\rho\alpha} \gamma^a_{\alpha\beta} \theta_{0\rho}^\beta \cr
\delta A &= \delta_0 A + 2\eta^{\rho\alpha}\theta_{0\rho\alpha} \cr
}}
on the other fields.
We have realized the algebra of the broken supersymmetry
$\tilde\eta^\rho_{\alpha}$ and unbroken $\eta^\rho_{\alpha}$ manifestly
on the parameters -- alternatively, somewhat simpler transformation laws
could be obtained by
rewriting the Lagrangian in terms of combinations $X-x_0$,
$\theta-\theta_0$ and $A-a_0$
invariant under the broken supersymmetry.
The unbroken supersymmetry
$\tilde\eta^{\drho}_{\alpha}$ does not act on the new sector.
The broken supersymmetries $\eta^{\drho}_{\alpha}$
should also be realized in some non-linear way, but making this
manifest appears to be more complicated.

By introducing auxiliary fields $D^{\rho\sigma}$,
the Lagrangian can be made quadratic in $v$.
This is an easy way to see that an overall coupling
constant can be absorbed by field redefinition.
The external fields
of the five-brane will be produced by a one-loop effect, and thus
there is no adjustable charge.

\newsec{Dynamics of zero-branes}

In this section we argue that the modifications to zero-brane
dynamics produced by \lagvec\ agree with predictions from supergravity.

Let us first consider the ground state, a configuration in which all
of the zero-branes are far from the five-brane,
whose position we now take to be $x^a_0=0$.  The modes $v$ and $\chi$
are all massive and sit in their ground states.  By supersymmetry, their
contributions to the vacuum energy cancel.

A graviton far from $x^a_0=0$ can be studied using the effective Lagrangian
produced by integrating out $v$ and $\chi$.
We first check that this effective Lagrangian reproduces the metric of
the five-brane \refs{\guven,\duff}
as felt by the bound state of N zero-branes.
The light-cone zero-brane Lagrangian is
\eqn\fivebmet{
\CL_0 = {1\over 2R}(D_t x_{||})^2 +
	{1\over 2R}\l(1+{B\over r^3}\r)(D_t x_\bot)^2
}
where the constant $B=\kappa_{11}^2 T_5/4\pi^2$,
$T_5=(\pi/2\kappa_{11}^4)^{1/3}$
in the conventions of \duff,
and $\kappa_{11}$ is determined by the parameters
of \lagmat\ in a way implicit in \BFSS.

The computation of the one-loop effective Lagrangian for
a single zero brane is the field theory limit of the 0--4-brane
computation carried out in \DKPS.  They reproduced the $d=10$ dimensional
reduction of this metric and showed that the long distance
limit agreed with the known string theory result and four-brane tension.
This is related to eleven-dimensional five-brane tension as
$T_4=R T_5$, while the gravitational couplings are related as
$R/\kappa_{11}^2=1/\kappa_{10}^2$, so the result
has the correct normalization.
Since the new degrees of freedom are vectors of $U(N)$,
the leading large distance, large $N$ contribution to $\tr \dot X^2$
will be independent of $N$, which combined with the scaling of \BFSS\ produces
the correct dependence on $p_{11}=N/R$.
Thus a graviton feels the long distance metric of a five-brane.

As in \DKPS, we would like to claim that this result is exact to all
orders in perturbation theory.
In the present case this is purely a question of the field theory defined
by the combined Lagrangian \lagmat\ and \lagvec.
It is the dimensional reduction
of $d=4$, $\CN=2$ supersymmetric gauge theory for which this
non-renormalization theorem is well-known, and we expect it
to hold after dimensional reduction as well.
We do not expect non-perturbative effects to
change the behavior at distances much greater than the 11D Planck scale.

Local excitations of the five-brane world-volume will necessarily
carry longitudinal momentum $P^{11}$ in the IMF, and thus must be
identified with threshold bound states of the zero-branes, localized
around $x_0^a$.  The full quantum Hilbert space of excitations will be
reproduced by the same scheme of block diagonal wave functions as \BFSS,
allowing a new type of block for each five-brane and each number $N\ge 1$
of zero-branes.

Reproducing the five-brane spectrum requires
a single tensor supermultiplet of bound states for each $N$,
and there is some evidence for this conjectured spectrum.
First, the same result is required for the eleven-dimensional
interpretation of strongly coupled type \IIa\ string theory.
This leaves the question of whether the bound states
are present in pure D-brane quantum mechanics.
This was shown for $N=1$ in \refs{\sethi,\DKPS}, and the fact that the
long range fields are the same as in the string theory makes the
general statement very plausible.

This system is in some ways simpler than the pure 0-brane bound state
dynamics considered in \BFSS\ and it would be quite interesting to
formulate some of their physical conjectures here, especially those
regarding Lorentz invariance.

\newsec{The Dirac condition and Berry's phase}

We now proceed to consider a membrane in this background.
This will be a configuration with non-commuting expectation values
for the longitudinal membrane coordinates, say $X^5$ and $X^6$:
\eqn\memconf{\eqalign{
X^5 = R_5 P&\qquad X^6 = R_6 Q \cr
[P,Q ] &= 2\pi i.
}}
The other expectation values $X^1\ldots X^4$ and $X^7\ldots X^9$
are fixed to c-numbers $x^1$, etc... to describe the membrane ground
state.

The membrane should feel the five-brane
magnetic field through the components
$C_{56\mu}$ and the integrated world-volume coupling
\eqn\memcouple{\eqalign{
\int C^{(3)} &= \int dt\ \p_t X^\mu A_\mu(X) \cr
A_\mu(X) &\equiv \int dX^5 dX^6\ C_{56\mu}(X).
}}
Integrating out the vector degrees of freedom \lagvec\ must produce
such a term in the effective action.
The potential $A_\mu(X)$ will not be single-valued,
so there must be an ambiguity in this procedure.
This must be associated in some way with the point $X=0$
where vector degrees of freedom become massless.
Furthermore, if the membrane charge is correct,
the magnetic flux $F=dA$ integrated over
an $S^2$ surrounding the five-brane
in the transverse dimensions $(X^7,X^8,X^9)$ will be quantized in the
minimal Dirac unit, $\int F = 2\pi$ \nepteit.

To see these effects, we may consider motion $X(t)$
with extremely slow time dependence.
The connection $A_\mu(X)$ is then Berry's connection \Berry,
\eqn\berry{
A_\mu(X) = \vev{X;0|{\p\over\p X^\mu}|X;0}
}
on the ground state wave function $\ket{X;0}$.

Only the fermions see the direction of $X^\mu$, so only they could
produce the effect.  Furthermore, the DNT argument can be made
with a hyperplane only if it is infinite, and should be independent of
any small fluctuations of the membrane.  This suggests that the effect is
due to fermion zero modes on the membrane.

In the limit of infinite membrane volume, we can realize the operators
$Q$ and $P$ in the Schr\"odinger representation, $Q=\sigma$ and
$P=2\pi i \p/\p\sigma$, and
turn the membrane theory into an effective two-dimensional field theory.
The fermionic Hamiltonian derived from \lagvec\ becomes
\eqn\fermham{
H = \int d\sigma\
\bar{\chi}^\drho
\l(\gamma_5 2\pi i R_5{\p/\p\sigma} + \gamma_6 R_6 \sigma
+ \vec\gamma \cdot \vec X\r) \chi_\drho,
}
where we use $\vec X = (X^7,X^8,X^9)$ to indicate the three dimensions
transverse to both branes.
At this point we drop the doublet index $\drho$ and the symplectic reality
condition, and work with a four component complex spinor.
We then decompose this into the tensor product of two $SO(3)$ spinors,
writing $\vec\gamma=-i\gamma_5\gamma_6\vec\tau$ in terms of Pauli matrices
$\vec\tau$.

Now we can see the origin of the Berry phase corresponding
to the five-brane magnetic field.
Consider the effect due to
a chiral (under $\gamma_5\gamma_6$) zero mode $\chi_0$ of $\chi$.
We will show shortly that all other contributions to the Berry phase
cancel for rigid motions of the membrane.
It has a two-component wave function and the Hamiltonian
\eqn\zeroham{
H = \bar\chi_0 \vec X \cdot \vec\tau \chi_0,
}
the same as for a spin $1/2$ particle in the magnetic field $\vec B=\vec X$.

As is well known, the Berry connection \berry\ for this system is exactly the
magnetic monopole with charge the Dirac quantum \Berry.
This is easy to verify by explicit computation of the ground state wave
functions, as is done in textbooks \naka.
The multi-valuedness of $A_\mu(X)$ arises because it is impossible to
choose a single phase convention for the ground state everywhere on
configuration space.  This is a consequence of the degeneracy of the ground
state at $X=0$.

Thus,
if we can show that membrane has a single chiral zero mode,
it follows that the membrane and five-brane are both
charged and satisfy the minimal Dirac condition.
The non-zero mode contributions are obtained by working in a basis of
eigenfunctions of
$H_0=\gamma_5 2\pi i R_5{\p/\p\sigma} + \gamma_6 R_6 \sigma$.
This is also very standard and the spectrum is $E^2_{n,\pm} = n$ with the
integer $n\ge 0$ for one chirality of $\gamma_5\gamma_6$ and $n \ge 1$
for the other chirality.  Since the two chiralities produce opposite
Berry phase, the non-zero mode contributions cancel, and only the single
chiral mode $E_{0,+}$ contributes.

We should check that the zero-brane does not feel the Berry phase.
Now the interaction
$\chi \gamma_5\gamma_6\vec X \cdot \vec\tau \chi$ is completely
symmetric under chirality reversal, so the phase completely cancels.
More generally, the monopole field appears only with three transverse
dimensions.

Similar fermionic zero modes will appear in a system containing
a dual pair of a D$p$-brane and a D$(6-p)$-brane.
Perhaps a variation of this argument can be found
to show that they satisfy the minimal Dirac condition, as found in \PolRR.

\newsec{Conclusions}

We proposed a modification of the M theory Lagrangian of \BFSS\ to
describe a longitudinal five-brane, checked that a zero-brane sees the
correct long-range fields, and checked that the membrane has
the correct Dirac unit of charge.  In combination these results also
imply that the membrane tension is correct, a point recently verified by
Shenker by a quite different argument \shenap.  Thus we have new
evidence for the model's consistency.

A number of further tests can be done \toappear.   The local fields
induced on the membrane world-volume can be computed.
It should be possible to construct a membrane ending on the five-brane.
It will also be interesting to interpret the Higgs branch
of the zero-brane theory with multiple five-branes.
The problem of constructing a `transverse' five-brane (with $x^{11}$
dependence) remains.

An unusual point of the construction is that our modified Lagrangian
involves additional dynamical variables.\footnote*{
This point was stressed to us by T. Banks, who also provided the resolution of
the following paradox.}
We believe that in generic situations
they do not correspond to additional physical degrees
of freedom.  An analogy can be drawn with the role of the
off-diagonal matrix components in the original construction.  In the
D-brane context, states in which these variables are excited are
states with physical stretched strings.  In the M theory context, there
should be no stretched string states before compactification.  The
resolution of this paradox is that these states have diverging energy
in the eleven dimensional limit.  This argument also applies
to excitations of the vectors.  It suggests that a modified
Lagrangian without additional variables might exist in the limit.

Let us make some comments on the Dirac condition.
A central theme of D-brane physics and its M theory analog is the equivalence
between `bulk' space-time interactions (exchange of closed strings between
D-branes; the supergravity interaction in M theory) and quantum interactions
of modes associated with pairs of branes (stretched open strings between
D-branes, or off-diagonal matrix components and vectors in M theory).

Here we see this equivalence in its simplest form.
The Dirac condition is one of the simplest and most general consequences
of the combination of gauge theory and quantum mechanics -- only the
topological structure of the gauge field enters.
On the other hand, Berry's phase is one of the simplest aspects of the
Born-Oppenheimer approximation in quantum mechanics, and is known
to reflect topological structure of the configuration space.  In particular,
a singular connection can appear only if the adiabatic approximation breaks
down, as it does at $\vec X=0$.  The simplest way this can happen is for
two eigenvalues of the Hamiltonian to become degenerate, as
in the present case.  The $U(1)$ bundle defined by the phase of the
wave function is embedded in a larger $SU(2)$ bundle at the origin, and there
is a strong analogy with the realization of the Dirac monopole as the
long distance field around the 't Hooft-Polyakov monopole.  This well-known
analogy becomes a physical equivalence in our problem: the monopole field
of the five-brane {\it is} the Berry's phase monopole, and just as for the
't Hooft-Polyakov monopole its singularity at the origin is regulated by
embedding it in an $SU(2)$ bundle, but now this is just a larger
subspace of the full quantum Hilbert space.

\medskip
We acknowledge valuable conversations with T. Banks, N. Seiberg,
S. Shenker and E. Witten.

\listrefs
\end